\documentclass[%
 reprint,
nofootinbib,
amsmath,amssymb,
aps, physrev,
prd,
]{revtex4-2}

\usepackage{physics}
\usepackage{graphicx}% Include figure files
\usepackage{dcolumn}% Align table columns on decimal point
\usepackage{bm}% bold math
\usepackage{hyperref}% add hypertext capabilities
\usepackage{booktabs} % for \toprule, \midrule, \bottomrule in tables
\usepackage{siunitx}
\usepackage{comment}
\usepackage{ulem}
\usepackage[dvipsnames]{xcolor}

\begin{document}

% \preprint{APS/123-QED}

\title{\textbf{Non-Gaussian matter fluctuations from primordial magnetic fields and galaxy luminosity function}}%

\author{Teppei Minoda}
\email{teppei.minoda@nao.ac.jp}
\affiliation{Department of Astronomy, Tsinghua University, Beijing 100084, China}
\affiliation{Ishigakijima Astronomical Observatory, National Astronomical Observatory of Japan,
National Institutes of Natural Sciences, 1024-1 Arakawa, Ishigaki, Okinawa, 907-0024, Japan}
\author{Yi Mao}
\email{ymao@tsinghua.edu.cn}
\affiliation{Department of Astronomy, Tsinghua University, Beijing 100084, China}
\author{Shohei Saga}
\email{shohei.saga@yukawa.kyoto-u.ac.jp}
\affiliation{Institute for Advanced Research, Nagoya University, Furo-cho Chikusa-ku, Nagoya 464-8601, Japan}
\affiliation{Kobayashi-Maskawa Institute for the Origin of Particles and the Universe, Nagoya University, Chikusa-ku, Nagoya, 464-8602, Japan}
\author{Hayato Shimabukuro}
\email{shimabukuro@ynu.edu.cn}
\affiliation{South-Western Institute for Astronomy Research (SWIFAR), Yunnan University, Kunming, Yunnan 650500, China}
\affiliation{Key Laboratory of Survey Science of Yunnan Province, Yunnan University, Kunming, Yunnan 650500, China}
\affiliation{Graduate School of Science, Division of Particle and Astrophysical Science, Nagoya University, Chikusa-Ku, Nagoya, 464-8602, Japan}

\date{\today}

\begin{abstract}
Primordial magnetic fields (PMFs) represent a promising candidate for the origin of large-scale magnetic fields observed in astrophysical environments. Through the Lorentz force, PMFs generate matter density perturbations, leaving distinct imprints on the formation of cosmic structures. While previous studies have primarily characterized these imprints via the matter power spectrum --- a two-point statistic --- we demonstrate that the one-point statistics of PMF-induced density fluctuations reveal an intrinsically non-Gaussian probability distribution function (PDF), even when the PMFs themselves are statistically Gaussian. In particular, we find that the high-density tail of the PDF is enhanced by over two orders of magnitude relative to the Gaussian case. Adopting the Press-Schechter formalism, we show that this non-Gaussianity significantly boosts the abundance of rare objects, especially massive halos and those at high redshifts. We further compute the ultraviolet luminosity function of galaxies at $z=6$, $12$, and $16$, and compare our predictions with recent James Webb Space Telescope (JWST) observations that have reported an excess of high-redshift galaxies at $z\sim 12$ -- $16$ relative to standard $\Lambda$CDM expectations. Intriguingly, in the limiting case where the total density field inherits this PDF, our results suggest that the non-Gaussian density fluctuations sourced by PMFs can explain the bright-end ($M_\mathrm{UV}\lesssim -18$) excess at $z\gtrsim 12$, without requiring additional astrophysical enhancements to star-formation efficiency.
\end{abstract}

\maketitle

%\tableofcontents

\section{Introduction}
Magnetic fields can be observed in various astrophysical objects \cite{2012SSRv..166....1R, 2013pss5.book..641B, 2017ARA&A..55..111H}. In particular, the origin and evolution of magnetic fields on large scales ($\gtrsim \mathrm{Mpc}$, e.g., magnetic fields in galaxy clusters \cite{1990ApJ...355...29K, 1991ApJ...379...80K, 1993ApJ...416..554T, 2001ApJ...547L.111C, 2002ARA&A..40..319C} and intergalactic magnetic fields \cite{2010Sci...328...73N, 2010MNRAS.406L..70T, 2010ApJ...722L..39A}) are an open question in both astrophysics and cosmology.
One promising scenario is the generation of magnetic fields in the early universe, which are called the primordial magnetic fields (PMFs; see \cite{2011PhR...505....1K, 2013A&ARv..21...62D, 2016RPPh...79g6901S} for detailed reviews). Generation mechanisms of PMFs are suggested from various physical processes, such as the inflation and the phase transition \cite{2011JCAP...03..037D, 2013PhRvD..87b3504B, 2013PhRvD..87h3007K, 2016JCAP...10..039A}.

The role of PMFs in structure formation has been investigated from multiple perspectives. Early studies pointed out that the Lorentz force sourced by PMFs can generate additional density fluctuations after recombination, leading to enhanced small-scale power (e.g., \cite{1978ApJ...224..337W, 1996ApJ...468...28K}). Subsequent works developed linear perturbation theory including magnetic sources and characterized the resulting matter power spectrum (e.g., \cite{2003JApA...24...51G, 2012PhRvD..86d3510S, 2023PhRvD.108b3521A}).
Recently, non-linear structure formation and its impact on observables from PMFs has been investigated \cite{2014JCAP...08..017S, 2023PhRvL.131w1002R}, particularly using cosmological magnetohydrodynamics (MHD) simulations \cite{2016MNRAS.456L..69M,2021MNRAS.500.5350V,2021MNRAS.507.1254K,2022ApJ...929..127M,2024JCAP...07..027R,2025JCAP...08..011R,2025PhRvL.135g1001P}.
While most studies with MHD simulations have applied the ideal MHD approximation so far, the energy dissipation of the magnetic field and its effect on the intergalactic medium (IGM) gas heating are suggested to be significant \cite{2005MNRAS.356..778S, 2014PhRvD..89j3522S, 2019MNRAS.488.2001M}.

These physical processes (enhancement of matter power spectrum and the heating of IGM) can significantly affect cosmological observables such as the cosmic microwave background \cite{2019PhRvL.123b1301J, 2022MNRAS.517.3916P, 2025PhLB..86539456U}, the Ly$\alpha$ forest \cite{2025PhRvL.135g1001P}, dwarf-galaxy abundance \cite{2024A&A...690A..59S}, intrinsic alignments \cite{2024PhRvD.109d3520S}, and 21-cm lines \cite{Shimabukuro2014SmallScale21cmForest,2019MNRAS.488.2001M, 2020MNRAS.498..918B, Shimabukuro2020Axion21cmForest,Shimabukuro2020ULDM21cmForest, 2024PhRvD.109b3518C, 2025JCAP...01..089B}

In addition, it has recently been suggested that the density fluctuations induced by PMFs in the early universe cause the inhomogeneous recombination, and PMFs could relieve the $H_0$ and $\sigma_8$ tension \cite{2020PhRvL.125r1302J, 2023arXiv230705475J, 2025JCAP...03..012J, 2025arXiv250616517S}. Future observations would achieve a further understanding of PMFs \cite{2022PhRvD.105f3537M, 2023JCAP...08..017K, 2023JCAP...09..035A, 2024JCAP...07..086P, 2026arXiv260407327B,2026arXiv260422703F,2026arXiv260505323W}.

Despite these efforts, a systematic exploration of non-Gaussian statistics of PMF-induced density fluctuations and their propagation into observables remains incomplete.
Physically, the Lorentz source term is quadratic in $\bm{B}$, so that even when each component of $\bm{B}$ is drawn from a Gaussian distribution, the source field follows a $\chi^2$-like distribution with a pronounced positive tail, which propagates into the linear matter density contrast $\delta$ through the linear perturbation equation. This intrinsic non-Gaussianity has so far received little quantitative attention in the context of structure-formation observables.
In this paper, we investigate the non-Gaussianity of PMF-induced density fluctuations and their impact on the halo mass function (HMF) and galaxy UV luminosity function (UVLF).

The remainder of this paper is organized as follows.
In Sec.~\ref{sec:methods}, we summarize the statistical properties of PMFs and the matter density fluctuations they induce through the Lorentz source term, including their two-point and one-point statistics.
In Sec.~\ref{sec:results}, we describe our numerical setup and present the resulting non-Gaussian one-point probability distribution function (PDF) of the PMF-sourced density fluctuations.
In Sec.~\ref{sec:HMF-UVLF}, we propagate these statistics through the Press-Schechter formalism to the HMF and the galaxy UVLF at $z=6$, $12$, and $16$.
We discuss the implications and limitations of our results and conclude in Sec.~\ref{sec:conclusion}. We leave some technical details to Appendix~\ref{app:astro-sensitivity} (regarding the sensitivity of the UVLF to the astrophysical parameters).
Throughout this paper, we adopt a flat $\Lambda$CDM cosmology with cosmological parameters $\Omega_\mathrm{m}=0.315$, $\Omega_\mathrm{b}=0.0493$, $h=0.674$, $\sigma_8=0.811$, $n_s=0.965$ \cite{2020A&A...641A...6P}.

\section{Basics of PMFs\label{sec:methods}}

\subsection{Statistical properties of PMFs}
First, we need to prepare the statistical properties of PMFs. In this work, PMFs are assumed to evolve adiabatically, $\bm{B}(\bm{x},z) = \bm{B}(\bm{x},0) (1+z)^{2}$. Hereafter, we treat the PMFs in the comoving units unless explicitly mentioned.

Similar to many previous works, we assume that the PMFs are statistically homogeneous and isotropic. Therefore, the spatial distribution of PMFs can be completely described by the power spectrum as
\begin{align}
\langle B_i(\bm{k})B^*_j(\bm{k}') \rangle = \cfrac{(2\pi)^3}{2} \delta^3_\mathrm{D} (\bm{k-k}') (\delta_{ij}-\hat{k}_i\hat{k}_j) P_B (k)~.
\end{align}
Here, $\hat{\bm{k}}$ is the normalized wave number vector, and $\bm{B}(\bm{k}) = \int \dd^3 x ~\mathrm{e}^{-i\bm{k}\cdot\bm{x}} \bm{B}(\bm{x})$ represents the magnetic fields in Fourier space. We also neglect the helicity of PMFs.

A widely used formulation of the PMF power spectrum $P_B(k)$ is the power-law type with a cutoff at smaller scales,
\begin{align}
  P_B(k) =
  \begin{cases}
  A_B k^{n_B} & k<k_\mathrm{cut} \\
  0           & k\ge k_\mathrm{cut}
  \end{cases}
  \label{eq: def P_B}  ~,
\end{align}
where $A_B$ and $n_B$ parametrize the amplitude and spectral index of the power spectrum. In this model, the PMF power spectrum has a cutoff scale, $k_\mathrm{cut}$, due to the radiative diffusion effect in the early universe \cite{1998PhRvD..57.3264J, 1998PhRvD..58h3502S}.
%On smaller scales than the cutoff, the PMF energy is overdamped; i.e.,
%\begin{align}
%P_B(k) = 0 \hspace{1cm} \text{(for $k\ge k_\mathrm{cut}$)}~.
%\label{eq: P_B k>}
%\end{align}

Instead of using the amplitude of the power spectrum $A_{B}$, we introduce the normalized magnetic field strength $B_\mathrm{n}$ as \cite{2019MNRAS.484..185P}
\begin{align}
B_\mathrm{n}^2
= \int \cfrac{\dd^3 k}{(2\pi)^3} ~P_B(k)
= \cfrac{A_B}{2\pi^2 (n_B+3)} k_\mathrm{cut}^{n_B+3} ~.
\end{align}
Based on the linear analysis of perturbed magnetic field lines, the cutoff scale is obtained by \cite{1998PhRvD..57.3264J, 1998PhRvD..58h3502S, 2002PhRvD..65l3004M, 2019MNRAS.484..185P}
\begin{align}
k_\mathrm{cut} =& 2.3 \times 10^2
\sqrt{\cfrac{(2 \pi)^{n_B+3}}{\Gamma[(n_B+5)/2]}}
 \nonumber \\
&\times \left(\cfrac{B_\mathrm{n}}{\mathrm{1~nG}}\right)^{-1}
h^{1/2}
\left(\cfrac{\Omega_\mathrm{b} h^2}{0.022}\right)^{1/2} ~\mathrm{[Mpc^{-1}]} ~.
\end{align}
As a result, we parametrize the PMFs by $B_\mathrm{n}$ and $n_B$.

\subsection{Matter density fluctuations from PMFs \label{sec:MatterPowerSpectrum}}
\label{sec:NonGaussianity}
It is pointed out that PMFs induce matter density fluctuations due to the Lorentz force
\cite{1978ApJ...224..337W}.
We organize our discussion of these fluctuations into three parts: the Lorentz source term, two-point statistics through the matter power spectrum and the mass dispersion $\sigma(M,z)$, and one-point statistics through the distribution $P_\mathrm{1pt}(\delta_M|M,z)$.

First, we briefly explain the evolution of the density fluctuations driven by the Lorentz source term.
We define the linear matter density contrast at cosmic time $t$ as
$\rho_\mathrm{m}(\bm{x},t) = \bar{\rho}_\mathrm{m}(t)\,[1+\delta(\bm{x},t)]$~,
where $\rho_\mathrm{m}$ is the matter density and $\bar{\rho}_\mathrm{m}$ is its mean.
Under the linear, pressure-free approximation, its evolution in the presence of PMFs is given by \cite{1978ApJ...224..337W}
\begin{align}
  \label{eq:delta_m}
  \Ddot{\delta}(\bm{x},t) + 2H(t) \Dot{\delta}(\bm{x},t)
  -4\pi G \rho_\mathrm{m}(t) \delta(\bm{x},t) \nonumber \\
  = \cfrac{\nabla \cdot \left[(\nabla \times \bm{B}(\bm{x},t)) \times \bm{B}(\bm{x},t)\right]}{4\pi \rho_\mathrm{m}(t)}~.
\end{align}
Here $\nabla$, $\bm{B}$, and $\rho_\mathrm{m}$ denote proper (physical) quantities.
The right-hand side is a quadratic form in $\bm{B}$, and therefore, even if $\bm{B}$ is a Gaussian random field, the source term follows a non-Gaussian (heavy-tailed) distribution.

Eq.~\eqref{eq:delta_m} is valid under two crucial assumptions: the pressure-free approximation and the linear approximation.
The pressure-free assumption is strictly valid only on scales much larger than the baryonic and magnetic Jeans scales, where thermal and magnetic pressure forces are subdominant compared with gravity \cite{1978ApJ...224..337W, 1996ApJ...468...28K}.
The magnetic Jeans scale is given by \cite{1996ApJ...468...28K,2012JCAP...11..055F,2019MNRAS.484..185P,2024PhRvD.109b3518C}
\begin{align}
\lambda_B
& =\left[
\frac{4}{25}
\frac{B_\mathrm{n}^2}{G \rho_\mathrm{m, 0} \rho_\mathrm{b, 0}}
\frac{(n_B+3)}{2\,k_{\rm cut}^{n_B+3}}\,
\Gamma\!\left(\frac{n_B+3}{2}\right)
\right]^{1/(n_B+5)} \nonumber
\\
& \simeq
\left[ 0.06\, (n_B+3) \,
\Gamma\!\left(\frac{n_B+3}{2}\right) \right. \nonumber
\\
& \hspace{1cm} \left.
\times
\left(\frac{B_n}{1\,\mathrm{nG}}\right)^{2}
\left(\frac{1\,\mathrm{Mpc}^{-1}}{k_{\rm cut}}\right)^{n_B+3}
\right]^{1/(n_B+5)} \nonumber
\\
& \hspace{62mm} \mathrm{[Mpc]}.
\end{align}
As we will discuss in Sec.~\ref{sec:NumericalSetup}, we take the simulation box size and numerical resolution to be large enough to satisfy this assumption.

Next, we introduce the two-point statistics of the PMF-induced matter density fluctuations.
These fluctuations contributes additively to the linear matter power spectrum.
The total matter power spectrum at the present epoch is simply written as
\begin{equation}
P_\mathrm{m}(k) = P_\mathrm{m}^{(\Lambda \mathrm{CDM})}(k) + P_\mathrm{m}^{(B)}(k)~.
\end{equation}
Here we neglected the correlation between matter density fluctuations predicted by the cosmic
inflation and ones induced by PMFs, for simplicity.
Besides, the redshift dependence of the matter power spectrum is assumed to be given by the linear growth factor as $P_\mathrm{m}(k,z)=P_\mathrm{m}(k) D^2(z)$, for simplicity.

\begin{figure*}
\centering
% Renamed to match the figure number in print (was Fig00_matterpower_1x3.pdf).
\includegraphics[width=1.0\linewidth]{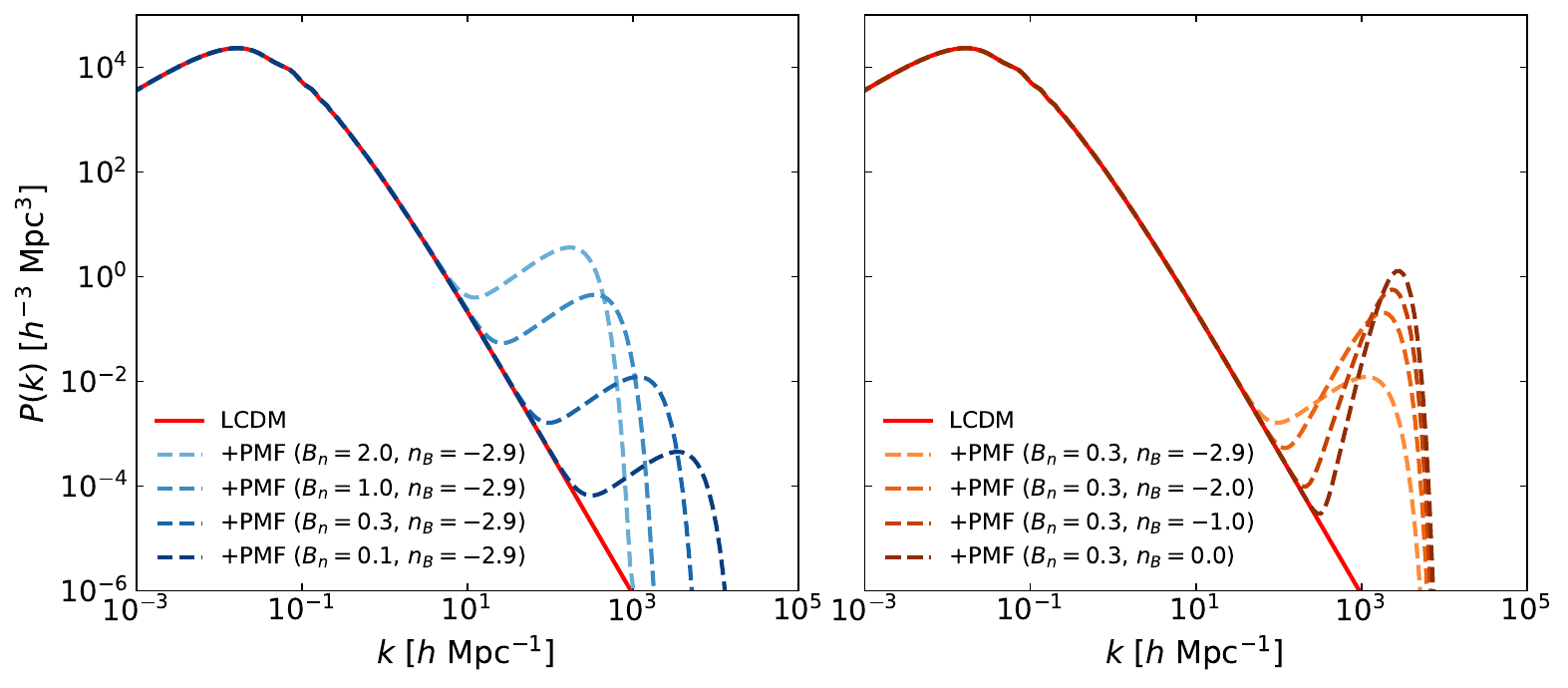}
\caption{\label{fig:PMFmatter}PMF-induced matter power spectra at $z=0$ with different values of $B_n$ (left) and $n_B$ (right).}
\end{figure*}

Ref. \cite{2021JCAP...11..044K} gives the approximated formula for the PMF-induced matter power spectrum as
\begin{align}
  &P_{\rm m}^{(B)}(k)=\frac{2 \pi}{k^3}\left(\frac{k}{a_0 H_0}\right)^4 \frac{4}{225}\left(1+z_{\mathrm{dec}}\right)^2
  \left(\frac{\Omega_{\gamma, 0}}{\Omega_{\rm m, 0}}\right)^2
  \left(\frac{\rho_{B 0}}{\rho_{\gamma 0}}\right)^2
  \notag \\
  & \times
  \left(\frac{k}{k_\mathrm{cut}}\right)^{2\left(n_B+3\right)}
  \frac{e^{-(k/k_{\rm cut})^{2}}}{\left[\Gamma\left(\frac{n_B+3}{2}\right)\right]^2}
  \sum_{i = 1}^{3}
  a_{i}\left(n_B+b_{i}\right)^{c_{i}n_{B}+d_{i}},
  \label{eq:P_m_b}
\end{align}
with the constants being $a_{1}=4.095$, $a_{2}=13.75$, $a_{3}=25.23$, $b_{1}=b_{3}=3.1$, $b_{2}=3.05$, $c_{1}=c_{3}=0$, $c_{2}=-0.330$, $d_{1}=-1.166$, $d_{2}=-2.158$, and $d_{3}=-0.993$.
Here, $z_\mathrm{dec} \simeq 1089$ denotes the redshift of photon decoupling at which the Lorentz force begins to source matter density fluctuations efficiently, and we use $\rho_{B 0}/\rho_{\gamma 0} = 9.545 \times 10^{-8} (B_\mathrm{n}/1 \text{nG})^2$ \cite{2021JCAP...11..044K}.
Based on Eq.~\eqref{eq:P_m_b}, we plot the PMF-induced matter power spectra for various values of the PMF parameters $B_\mathrm{n}$ and $n_B$ in Figure~\ref{fig:PMFmatter}.
Larger $B_\mathrm{n}$ makes the amplitude and the peak scale larger, and larger $n_B$ makes the power spectrum steeper.

\begin{figure*}
\centering
% Renamed to match the figure number in print (was Fig00-2_sigmaMrelation_1x3.pdf).
\includegraphics[width=1.0\linewidth]{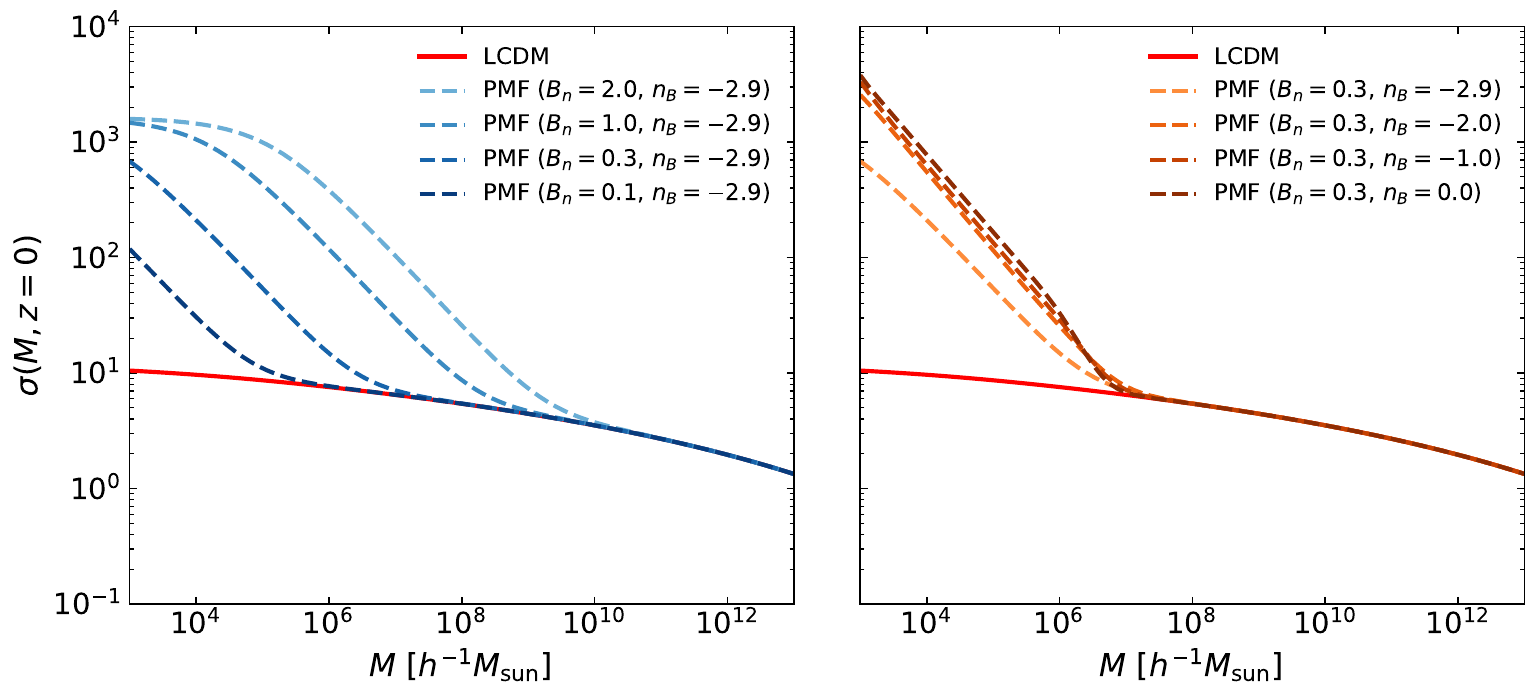}
\caption{\label{fig:MsigmaR}
Mass dispersion $\sigma(M,z=0)$ as a function of halo mass $M$ for different $B_n$ (left) and $n_B$ (right).
}
\end{figure*}

Matter power spectrum can be translated to the mass dispersion $\sigma(M,z)$, the amplitude of density fluctuations averaged with a given mass scale $M$ and redshift $z$. The mass dispersion is defined as $\sigma^2(M,z) \equiv \langle \delta_M^2 (\bm{x}, z) \rangle$, where $\delta_M (\bm{x}, z)$ is the density fluctuation smoothed on the length scale corresponding to a given mass $M$.
The relation between the matter power spectrum and the mass dispersion is given by
\begin{align}
  \sigma^2 (M,z) &= \int \cfrac{\dd^3 k}{(2\pi)^3} W^2(kR) P_\mathrm{m}(k,z) .
\end{align}
Here we have used the top-hat window function
$W(x) = \left( 3 \sin{x}-3x\cos{x} \right)/x^3$, and the scale mass $M$ corresponds to the filtering scale $R$ as $M = (4\pi/3) \rho_\mathrm{m,0} R^3$
\cite{1974ApJ...187..425P, 1986ApJ...304...15B}. Similarly to the matter power
spectrum in Figure \ref{fig:PMFmatter}, we also show the mass dispersion
including the effects of PMFs at $z=0$ in Figure \ref{fig:MsigmaR}.
The enhancement of the small-scale matter power spectrum increases $\sigma(M,z)$ at low filtering masses, $M \lesssim 10^{10}\,h^{-1}\,M_\odot$ for $B_\mathrm{n}=2.0$~nG and $M \lesssim 10^{5}\,h^{-1}\,M_\odot$ for $B_\mathrm{n}=0.1$~nG.
Although the PMF spectral index $n_B$ does not drastically change the $M$--$\sigma$ relation, it affects the steepness of enhanced mass-dispersion relation.
% \com{$M$ と $M_\mathrm{h}$ の使い分けをどう考えるか。論文の構成も合わせて考える。}

Having specified the two-point statistics through $\sigma(M,z)$, we construct the one-point distribution $P_\mathrm{1pt}(\delta_M|M,z)$.
While $\delta_M(\bm{x},z)$ denotes the smoothed spatial field, hereafter $\delta_M$ (i.e.\ without arguments) denotes the value of this field at a randomly chosen position, regarded as a random variable. Its one-point statistics is estimated from the spatial histogram of the field under the assumption of ergodicity.
To calculate the halo number count, in addition to the halo mass-dispersion relation, we need the one-point PDF of $\delta_M$, denoted as $P_\mathrm{1pt}(\delta_M)$ that gives the probability density for the scalar realization~$\delta_M$. In $\Lambda$CDM cosmology, the density fluctuations are often assumed to follow a Gaussian distribution,
\begin{align}
P_\mathrm{1pt}^{\text{(Gauss)}}(\delta_M | M,z) = \cfrac{1}{\sqrt{2 \pi \sigma^2(M,z)}} \exp \left(-\cfrac{\delta_M^2}{2\sigma^2(M,z)}\right) .
\label{eq:1ptpdf-gauss}
\end{align}
However, this assumption is no longer valid in presence of PMFs.
As long as the linear approximation is valid, since the solution of Eq.~\eqref{eq:delta_m} is time-space-separable as $\delta_M(\bm{x},z) = D(z)\, \delta_M^\mathrm{(lin)}(\bm{x},0)$, the time evolution of one-point PDF of $\delta_M$ merely rescales the overall dispersion without altering its shape.
Additionally, the spatial profile of $\delta_M^\mathrm{(lin)}(\bm{x},0) \propto \nabla \cdot [(\nabla \times \bm{B}) \times \bm{B}]$ depends only on the realisation of the magnetic field.
Consequently, the non-Gaussian shape of $P_\mathrm{1pt}(\delta_M)$ is fully determined by the spatial histogram of $\nabla \cdot [(\nabla \times \bm{B}) \times \bm{B}]$ evaluated at a single snapshot, independent of redshift. The redshift dependence enters only through the overall normalisation $\sigma(M,z)$.
In Sec.~\ref{sec:NumericalSetup}, we show how to explicitly calculate the spatial distribution of the Lorentz force term.

\section{Numerical implementation and one-point statistics}
\label{sec:results}

\subsection{Numerical setup}
\label{sec:NumericalSetup}
In this paper, we numerically calculate the non-Gaussianity of density fluctuations produced by PMFs. We prepare the calculation box with the grid number $N_\mathrm{grid}=64^3$ and the box size $L_\mathrm{box} = 500~\mathrm{Mpc}$. Therefore the grid size is $\Delta x \equiv L_\mathrm{box}/N_\mathrm{grid}^{1/3} = 7.8~\mathrm{Mpc}$. To investigate relations between statistical properties of PMFs and non-Gaussianity of PMF-induced density fluctuations, we generate realizations of the magnetic field in the simulation box and evaluate the PMF-sourced linear density contrast $\delta(\bm{x})$ on the grid for each PMF model characterized by $(B_n, n_B)$, following the source term in Eq.~\eqref{eq:delta_m}.
To satisfy the divergence-free condition $\bm{k}\cdot\bm{B}(\bm{k})=0$, we generate the vector field in Fourier space $\bm{A}(\bm{k})$ as a random Gaussian field and take rotation to obtain the PMFs $\bm{B(k)}=i\bm{k}\times\bm{A}(\bm{k})$. The variance of the vector field is set by the power spectrum $P_B(k)$ in Eq.~(\ref{eq: def P_B}).
Then we take outer product of $\nabla \times \bm{B}$ and $\bm{B}$ to obtain the Lorentz force term $\nabla \left[\cdot (\nabla \times \bm{B}) \times \bm{B}\right]$.

In this work we explore a range of PMF parameters, $B_n = \{0.1, 0.3, 1.0, 2.0\}~\mathrm{nG}$ and $n_B = \{-2.9, -2.0, -1.0, 0.0\}$, and perform 50 independent realizations for each parameter set in order to reduce the sample variance (see Table~\ref{tab:fitting-parameters}).
From these realizations, we construct the one-point histogram of the Lorentz source term on the grid; under the linear approximation it is equivalent to the histogram of PMF-sourced $\delta$ up to an overall normalization absorbed into $\sigma(M,z)$.
We choose the box size and the cell size to be large enough to satisfy the pressure-free assumption, i.e., $\Delta x \gg \lambda_{B}$.
We have verified that the shape of the normalized one-point PDF is insensitive to the smoothing scale as long as it is sufficiently larger than $k_\mathrm{cut}^{-1}$; as the smoothing scale approaches $k_\mathrm{cut}^{-1}$ the non-Gaussianity becomes significantly stronger, but such small scales lie below the magnetic Jeans scale $\lambda_B$, where magnetic pressure prevents density fluctuations from growing and hence haloes from forming.

\subsection{Non-Gaussianity from PMFs}
\label{sec:results-pdf}
\begin{figure*}
\centering
% Renamed to match the figure number in print (was Fig01_PDF_Deviation_2x3.pdf).
\includegraphics[width=1.0\linewidth]{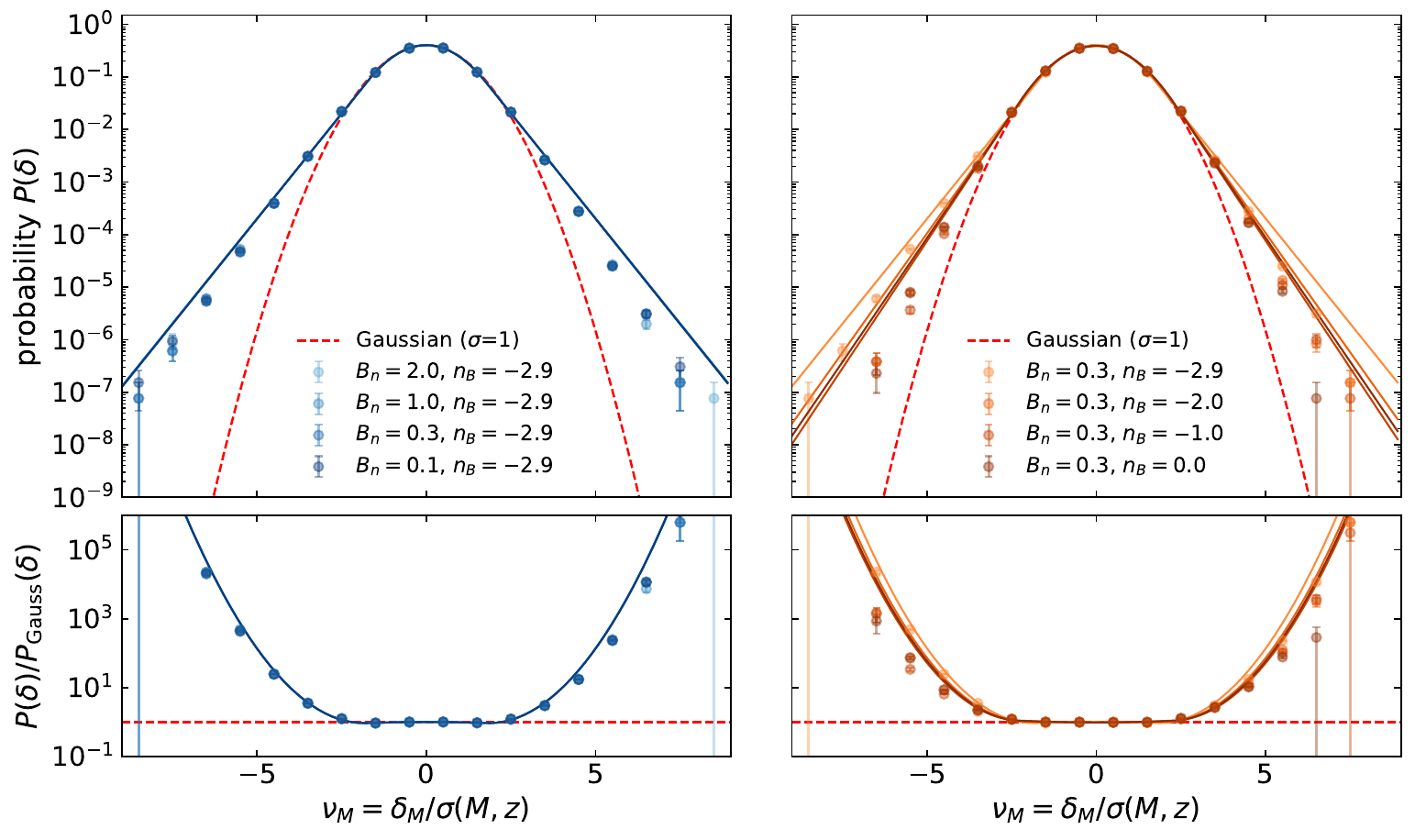}
\caption{\label{fig:PDF_delta_PMFs} One-point PDF of PMF-induced density fluctuations with different values of $B_n$ (left) and $n_B$ (right) plotted with $1\sigma$ error. Error bars on the datapoints are estimated as Poisson noise. The Gaussian PDF (red line) is also plotted for comparison, and the PMF-induced PDFs for different PMF parameters are shown by various colored lines. The bottom panel shows the probability ratio of PMF-induced density fluctuations to the Gaussian ones.}
\end{figure*}

In Figure \ref{fig:PDF_delta_PMFs}, we show the numerical results of the one-point PDF of PMF-induced density fluctuations. Here we do not include the adiabatic perturbations predicted by the inflation mechanism.
As shown in Sec.~\ref{sec:NonGaussianity}, within the linear approximation, the shape of the one-point PDF is independent of redshift once $\delta_M$ is normalized by $\sigma(M,z)$.
Therefore we plot the one-point PDF $P_\mathrm{1pt}$ as a function of the normalized variable $\nu_M \equiv \delta_M/\sigma$.
In the central region of probability, $|\delta_M| \lesssim 2 \sigma$, the deviation from Gaussian is smaller than $20\%$. However, in the outer region, $|\delta_M| \gtrsim 3 \sigma$,
the probability ratio $P_\mathrm{1pt}^{(\mathrm{PMF})} / P_\mathrm{1pt}^{(\mathrm{Gauss})}$ is no longer of order unity; for example, $P_\mathrm{1pt}^{(\mathrm{PMF})} / P_\mathrm{1pt}^{(\mathrm{Gauss})} \sim 10^3$ at $\delta_M \sim 6 \sigma$. When fixing $n_B$ and varying the PMF strength, the shape of the normalised one-point PDF is nearly universal. On the other hand, varying the spectral index of the PMF causes the tails of the PDF to change slightly depending on $n_B$. In particular, when $n_B = -2.9$, a significant deviation from a Gaussian PDF is produced.

Now we consider the physical implications for this near-universality.
For $n_B \gtrsim -2$, the induced density fluctuations are dominated by a narrow range of scales in Fourier space as shown in Fig. \ref{fig:PMFmatter}, close to the cutoff scale $k_\mathrm{cut}$.
For $n_B = -2.9$, in contrast, the PMF contributes to a wider range of spatial scales, and the mode-mixing effect can be significant.

As seen above, PMFs with a power-law spectrum produce an exponential tail in the one-point PDF of
density fluctuations; a similar feature has been reported for $\delta$-function-like PMFs
\cite{2020JCAP...05..039S}. An analytic expression for $P_\mathrm{1pt}$ would be desirable in order
to generalise the parameter dependence found numerically---in particular the near universality under
changes of $B_n$ and the milder sensitivity to $n_B$. Such a derivation is difficult, because
the Lorentz source $\nabla\cdot[(\nabla\times\bm{B})\times\bm{B}]$ is a quadratic functional of the
Gaussian field $\bm{B}$, whose one-point PDF is difficult to obtain in closed form. Instead of obtaining
the analytical solution of the one-point PDF of density fluctuations, we use the fitting function of
the numerical calculation results. The numerical results are fitted well by the exponential
function $P_\mathrm{1pt} \propto \exp{(-|\delta_M|/\sigma)}$ in the outside of the transition points,
$|\delta_M|>\delta_\text{tr}$. Imposing continuity and smoothness conditions, we obtain the fitting
function of 1-point PDF as
\begin{align}
P_\mathrm{1pt}^{\text{(fitting)}}(\nu_M) =
\left\{
\begin{array}{llll}
\cfrac{A}{\sqrt{2 \pi \sigma_\nu^2}}
\exp \left(-\cfrac{\nu_M^2}{2\sigma_\nu^2}\right) \\
\hspace{35mm} \text{, for}~ |\nu_M| \le \nu_\text{tr},\\
\cfrac{A}{\sqrt{2 \pi \sigma_\nu^2}}
\exp \left(-\cfrac{\nu_\text{tr}\left(|\nu_M|-\nu_\text{tr}\right)}{\sigma_\nu^2}
-\cfrac{\nu_\text{tr}^2}{2\sigma_\nu^2}\right) \\
\hspace{35mm} \text{, for}~ |\nu_M|>\nu_\text{tr}.
\end{array}
\right.
\label{eq:1ptpdf-exp}
\end{align}
with $\nu_M \equiv \delta_M/\sigma(M,z)$ and
$\nu_\text{tr} \equiv \delta_\text{tr}/\sigma(M,z)$. 
Here $\nu_\text{tr}$ and
$\sigma_\nu$ are fitting parameters, and the PDF is normalized with $A$ as
$\int_{-\infty}^{\infty} P_\mathrm{1pt}^{\text{(fitting)}}(\nu_M)\, \dd \nu_M=1$.
Note that $\delta_\text{tr}$ and $\nu_\text{tr}$ are chosen to be positive. The parameter $\sigma_\nu$ would be unity for the ideal Gaussian case, but for our non-Gaussian scenario, $\sigma_\nu$ is not unity for a fitting with a finite transition point $\nu_{\rm tr}$. 
Based on these numerical results, we determine the best-fit values of the transition
parameter $\nu_\text{tr}$ and the width parameter $\sigma_\nu$ by fitting the
simulated one-point PDF with Eq.\eqref{eq:1ptpdf-exp}. PMF model parameters and
derived best-fitted parameters in Eq.\eqref{eq:1ptpdf-exp} are summarized in
Tab. \ref{tab:fitting-parameters}.

\begin{table}
  \setlength{\tabcolsep}{8pt}
  \renewcommand{\arraystretch}{1.1}
  \centering
  \caption{Numerical setup parameters of calculation of density fluctuations induced by PMFs, together with the fitting parameters of 1-point PDF of PMF-induced density fluctuations. Our fiducial model is Model 3 (the same as Model 5).}
  \label{tab:fitting-parameters}
  \begin{tabular}{cccccc}
      \hline\hline
       model & $B_n$ [nG] & $n_B$ & $\lambda_{B}$ [Mpc] & $\nu_\text{tr}$ & $\sigma_\nu$ \\
      \hline
      1 & 2.0 & -2.9 & 0.96 & 1.76 & 0.98 \\
      2 & 1.0 & -2.9 & 0.50 & 1.76 & 0.98 \\
      3 & 0.3 & -2.9 & 0.16 & 1.76 & 0.98 \\
      4 & 0.1 & -2.9 & 0.055 & 1.77 & 0.98 \\
      5 & 0.3 & -2.9 & 0.16 & 1.76 & 0.98 \\
      6 & 0.3 & -2.0 & 0.27 & 2.16 & 1.02 \\
      7 & 0.3 & -1.0 & 0.38 & 2.35 & 1.03 \\
      8 & 0.3 &  0.0 & 0.46 & 2.27 & 1.02 \\
      \hline\hline
  \end{tabular}
\end{table}

\section{Implication for halo and galaxy abundance}
\label{sec:HMF-UVLF}

\subsection{The Impact of PMFs on the HMF}
\label{sec:results-HMF}

\begin{figure*}
  \centering
  % Renamed to match the figure number in print (was Fig00-3_HMF_2x2.pdf).
  \includegraphics[width=1.0\linewidth]{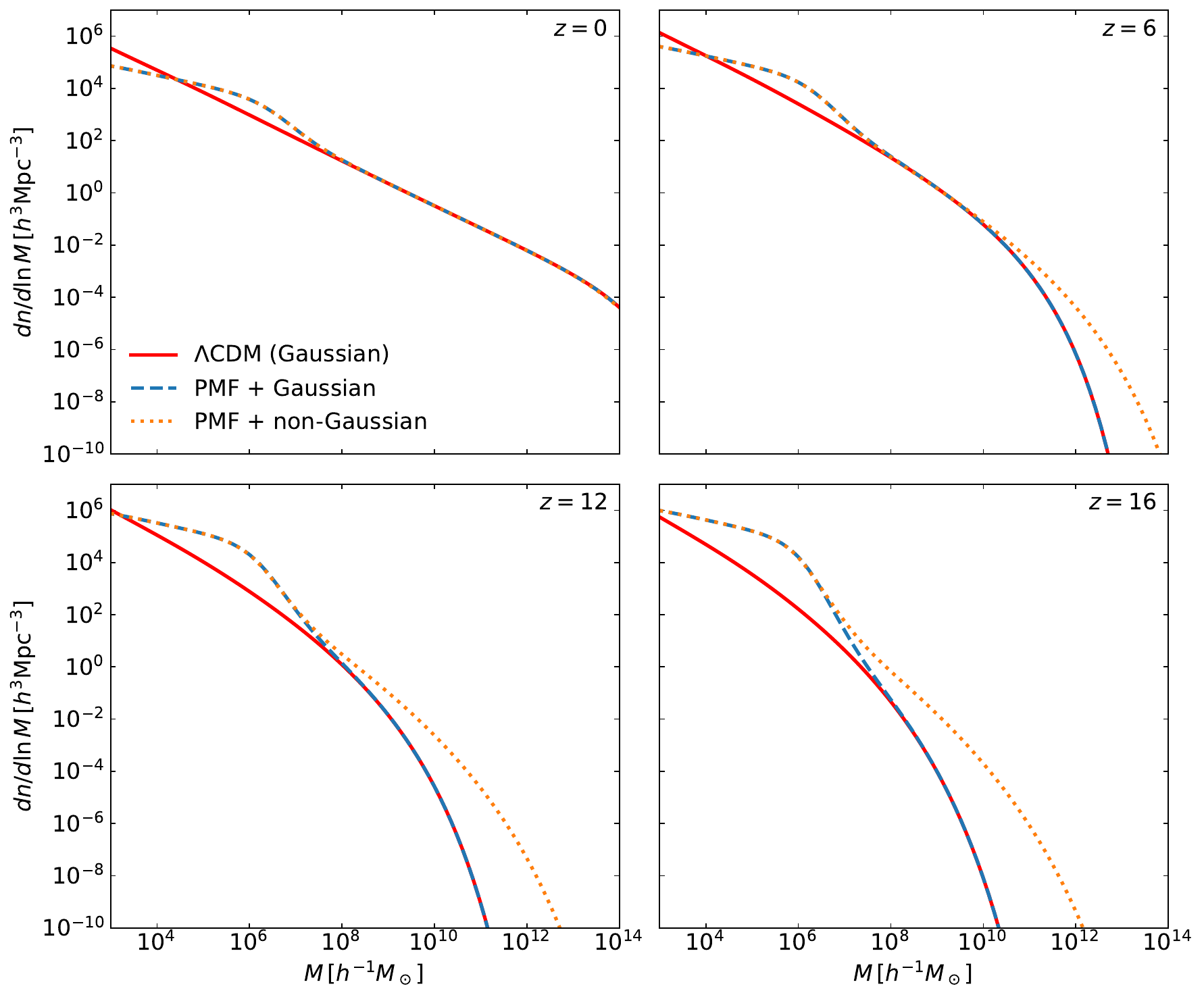}
  \caption{\label{fig:HMF} HMF at different redshifts of $\Lambda$CDM (solid), PMF-induced matter power spectrum with Gaussian PDF (dashed), and that with non-Gaussian PDF (dotted). In this figure, we show the results for Model 3 ($B_n=0.3$ nG, $n_B=-2.9$). The corresponding UVLFs, computed with exactly the same three prescriptions, are shown in Fig.~\ref{fig:LF_NGcom}.}
\end{figure*}

In this work, we calculate the HMF, incorporating the PMF-induced matter power spectrum and non-Gaussianity within the Press--Schechter framework.
As a limiting case, we assume that the total density field follows the non-Gaussian PDF shape of Eq.~(11), with its variance $\sigma(M,z)$ computed from Eq.~(7) (see the discussion in Sec.~IV B).
The comoving HMF is expressed in terms of the one-point PDF $P_\mathrm{1pt}(\delta_M)$ introduced in Sec.~\ref{sec:NonGaussianity} as
\begin{align}
\cfrac{\dd n}{\dd M} (M,z) = -\cfrac{2\bar{\rho}_\mathrm{m,0}}{M}
\cfrac{\partial}{\partial M} \int^\infty_{\delta_c} P_\mathrm{1pt}(\delta_M | M,z)~\dd \delta_M~.
\end{align}
In realistic flat $\Lambda$CDM models, the redshift evolution (and cosmology-dependence) of $\delta_c$ is well known to be negligible \cite{1980lssu.book.....P,1993MNRAS.262..627L}.
In this work, for simplicity, we adopt a constant threshold $\delta_c=1.69$\footnote{
Ref.~\cite{2014JCAP...08..017S} analyzed the effect of PMFs on the spherical-collapse model and showed that the critical linear overdensity for collapse can be significantly larger than the $\Lambda$CDM value particularly at high redshifts, which would reduce high-$z$ halo abundances relative to models with fixed $\delta_c$.}; we comment further on this approximation in Sec.~\ref{sec:conclusion}.
Therefore, the redshift dependence of $\dd n/\dd M$ enters only through $\sigma(M,z)$ in $P_\mathrm{1pt}(\delta_M | M,z)$.

We show the effect of PMFs on the Press-Schechter HMF in Fig. \ref{fig:HMF} at $z=0$, 6, 12, and 16. Here, we show the case of PMFs for Model 3 ($B_n=0.3$ nG, $n_B=-2.9$). The effect of PMF-induced matter power spectrum on mass function appears on $M \lesssim 10^7 ~h^{-1} ~M_\odot$, which corresponds to the results of the mass-dispersion relation. For very small-mass haloes ($M \lesssim 10^5 ~h^{-1} ~M_\odot$) at low redshifts ($z=0$ and 6), the number density of dark matter haloes is reduced by PMFs, which can be caused by the hierarchical structure formation.

In Fig. \ref{fig:HMF}, we compare two different calculation results including PMFs: (i) PMF-induced matter power spectrum and assuming the Gaussian one-point PDF (dashed lines), and (ii) PMF-induced matter power spectrum and assuming non-Gaussian PDF obtained in Sec. \ref{sec:results-pdf}. Non-Gaussian features significantly enhance the HMF for high-redshift and massive haloes; $M \gtrsim 10^{10} ~h^{-1} M_\odot$ for $z=6$, and $M \gtrsim 10^{7} ~h^{-1} M_\odot$ for $z=12$ and 16. These haloes are formed by large-scale density fluctuations with small $\sigma(M,z)$, and thus the abundance of such high-$z$ and massive objects is determined by the tail of $P_\mathrm{1pt}(\nu_M)$, where the PMF-induced PDF exceeds the Gaussian one by orders of magnitude as shown in Fig.~\ref{fig:PDF_delta_PMFs}.
For these rare objects, the collapse threshold $\delta_c/\sigma(M,z)$ lies beyond the transition point $\nu_\mathrm{tr}$ in Eq.~\eqref{eq:1ptpdf-exp}, so that the collapsed fraction is dominated by the exponential tail rather than by the Gaussian core of the PDF.
Since the exponential tail decays much more slowly than the Gaussian one, the relative enhancement of the HMF grows with increasing halo mass and redshift, which explains why the non-Gaussian effect extends down to smaller halo masses at higher redshifts in Fig.~\ref{fig:HMF}.

\subsection{The Impact of PMFs on the UVLF}

Following Ref.~\cite{2019MNRAS.484..933P}, we calculate the UVLF using the same matter power spectrum and non-Gaussian $P_\mathrm{1pt}(\delta_M)$ as in the halo calculation above.
The UVLF is given by \cite{2019MNRAS.484..933P}
\begin{align}
\phi(M_{\rm UV}) = f_{\rm duty} \cfrac{\dd n}{\dd M}
\left|\cfrac{\dd M}{\dd M_{\rm UV}}\right|~,
\label{eq:UVLF}
\end{align}
where $f_{\rm duty} = \exp(-M_\mathrm{turn}/M)$ is duty cycle of galaxies, $M$ is halo mass, $M_\mathrm{turn}$ is turnover mass, and $M_{\rm UV}$ is the galaxy UV magnitude.

The halo mass to UV magnitude relation, $\dd M/\dd M_{\rm UV}$ is obtained as follows.
The stellar mass-to-halo mass ratio is defined as
\begin{align}
\cfrac{M_*}{M} = f_*\cfrac{\Omega_{\rm b}}{\Omega_{\rm m}}~,
\label{eq:SMtoHM}
\end{align}
with the fraction of stars in baryon gas in a galaxy $f_*$, which is assumed as a single power-law of halo mass as
\begin{align}
f_* = f_{*,10} \left(\cfrac{M}{10^{10}M_\odot }\right)^{\alpha_*}~.
\end{align}
The galactic stellar mass determines the star formation rate as
\begin{align}
\dot M_{*}(z) = \cfrac{M_*}{t_* H^{-1}(z)}~.
\end{align}
Here $t_*$ is a dimensionless parameter, assumed to be independent of redshift, which characterizes the star-formation timescale in units of the Hubble time, so that the star formation rate is inversely proportional to the Hubble time, i.e., $\dot M_* \propto H(z)$.
The star formation rate is converted to the UV luminosity as \cite{2016MNRAS.460..417S}
\begin{align}
L_{\rm UV} = \cfrac{\dot M_{*}(z)}{1.15 \times 10^{-28} ~M_\odot \mathrm{~yr^{-1}}}
~[\mathrm{erg ~s^{-1} ~Hz^{-1}}]~,
\end{align}
and the UV luminosity is translated into the UV magnitude as \cite{1983ApJ...266..713O}
\begin{align}
M_{\rm UV} = 51.63 - 2.5\log_{10} \left(\cfrac{L_{\rm UV}}{\mathrm{erg ~s^{-1} ~Hz^{-1}}}\right)~.
\label{eq:UVmag}
\end{align}
Combining Eqs.\eqref{eq:SMtoHM}-\eqref{eq:UVmag}, we derive $\dd M/\dd M_{\rm UV}$ in Eq.\eqref{eq:UVLF}.
The UVLF therefore depends on four phenomenological parameters: the normalization $f_{*,10}$ and slope $\alpha_*$ of the stellar-to-halo mass ratio, the turnover mass $M_\mathrm{turn}$, and the star-formation timescale $t_*$.
We adopt the fiducial values of $f_{*,10}=0.05$, $\alpha_*=0.5$, $M_\mathrm{turn}=5 \times 10^{8} M_\odot$, and $t_*=0.5$ as in Ref.~\cite{2019MNRAS.484..933P}.
In the main text we keep these fiducial values fixed in order to isolate the impact of PMFs on the UVLF; the sensitivity of the UVLF to variations of the astrophysical parameters is examined in Appendix~\ref{app:astro-sensitivity}.

\begin{figure*}
\centering
\includegraphics[width=1.0\linewidth]{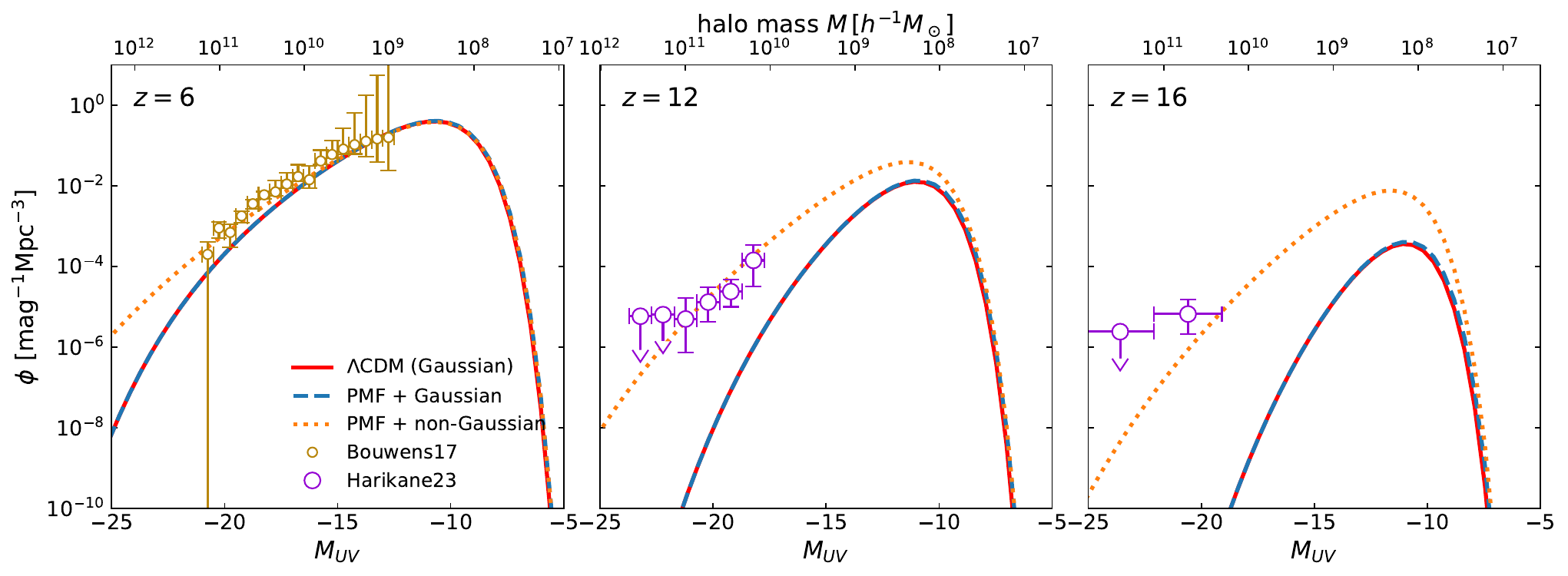}
\caption{\label{fig:LF_NGcom}
UVLF at $z=6$, $12$, and $16$ (left to right) including the PMF-induced density fluctuations. Here we adopt the fiducial parameters for PMFs (Model 3 in Tab.~\ref{tab:fitting-parameters}, $B_n = 0.3$~nG, $n_B = -2.9$) and astrophysical parameters. Red solid: $\Lambda$CDM with the Gaussian PDF. Blue dashed: PMF-induced matter power spectrum with the Gaussian PDF. Orange dotted: the same power spectrum together with the non-Gaussian PDF of Eq.~\eqref{eq:1ptpdf-exp}. The upper horizontal axis gives the halo mass corresponding to each $M_\mathrm{UV}$, which depends on redshift. 
For comparison, we plot the observational data points taken from Ref.~\cite{2017ApJ...843..129B} $(z=6)$ and Ref.~\cite{2023ApJS..265....5H} $(z=12$ and $16)$. 
}
\end{figure*}

PMFs affect the UVLF via two channels as same as they do the HMF: the additive contribution to the matter power spectrum, particularly on small scales, and the non-Gaussian shape of $P_\mathrm{1pt}$, which populates the far tail of the density distribution. Figure~\ref{fig:LF_NGcom} separates the two for the fiducial Model 3 ($B_n=0.3$~nG, $n_B=-2.9$), as Fig.~\ref{fig:HMF} does for the HMF. The upper horizontal axis of each panel converts $M_\mathrm{UV}$ into the halo mass $M$ that hosts galaxies of that magnitude, through Eqs.~\eqref{eq:SMtoHM}--\eqref{eq:UVmag}; since the star-formation rate scales with $H(z)$, this mapping is redshift dependent, and a given $M_\mathrm{UV}$ corresponds to a different halo mass in each panel.

As Fig.~\ref{fig:HMF} shows, the two channels act on different ranges of halo mass. The non-Gaussian tail of $P_\mathrm{1pt}$ boosts the abundance of rare massive haloes, $M\gtrsim 10^{10}\,h^{-1}\,M_\odot$ at $z=6$ and $M\gtrsim 10^{7}\,h^{-1}\,M_\odot$ at $z=12$ and $16$. The enhanced matter power spectrum, in contrast, changes the HMF only for small haloes, $M\lesssim 10^{8}\,h^{-1}\,M_\odot$. Since halo mass maps onto UV magnitude, the two effects should also appear at different galaxy luminosities in the UVLF.

Figure~\ref{fig:LF_NGcom} confirms this picture at the bright end, $M_\mathrm{UV}\lesssim -15$, which is populated by haloes with $M\gtrsim 10^{9}\,h^{-1}\,M_\odot$ and is the range probed by observations. Here the non-Gaussian PDF has a strong effect: it lifts the UVLF by more than two orders of magnitude at $z=12$ and $16$, which brings it into agreement with the JWST measurements, while the $z=6$ UVLF stays consistent with the data. The enhanced matter power spectrum, on the other hand, has only a marginal effect at the bright end, as expected from the HMF in Fig.~\ref{fig:HMF}.

At the faint end, however, neither effect shows up in the UVLF. In particular, the PMF$+$Gaussian curve cannot be distinguished from the $\Lambda$CDM one over the whole range plotted, down to $M_\mathrm{UV}=-5$, even though the underlying HMF differs by orders of magnitude at the corresponding masses. The cause is the duty cycle. The turnover halo mass $M_\mathrm{turn}=5\times 10^{8}\,M_\odot$ corresponds to $M_\mathrm{UV}\simeq -13$, and $f_\mathrm{duty}$ suppresses star formation in haloes below this mass. The haloes where the extra small-scale power matters lie well below $M_\mathrm{turn}$, so they are suppressed before they can add any UV luminosity. For the same reason, the part of the non-Gaussian enhancement that reaches below $M_\mathrm{turn}$ at $z=12$ and $16$ leaves no imprint on the faint-end UVLF either.

\begin{figure*}
\centering
% Renamed to match the figure number in print (was Fig06_LF_3z_nm29.pdf).
\includegraphics[width=1.0\linewidth]{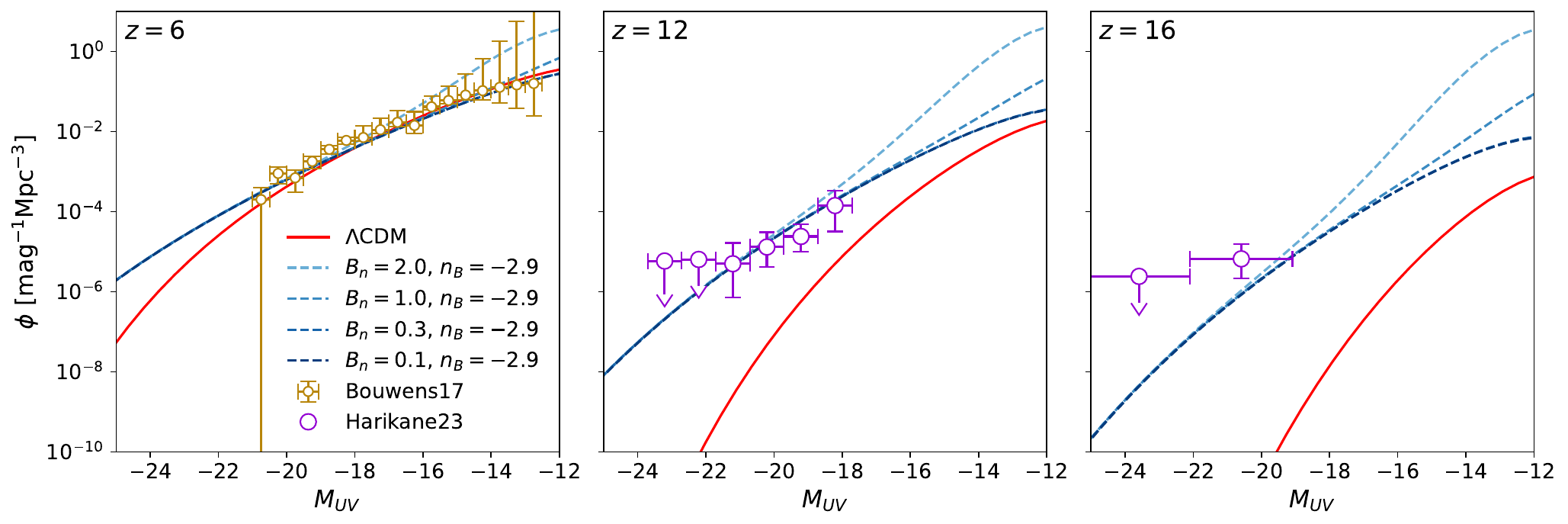}
\caption{\label{fig:LF_nm29} UVLFs at different redshifts of $\Lambda$CDM (solid) and PMF (dashed), the latter including both the PMF-induced matter power spectrum and the non-Gaussian $P_\mathrm{1pt}$.
From left to right, the corresponding redshifts are $z=6, 12 , 16$. 
Here, we show the PMF results for Model 1--4; darker shades correspond to weaker fields.}
\end{figure*}

\begin{figure*}
\centering
\includegraphics[width=1.0\linewidth]{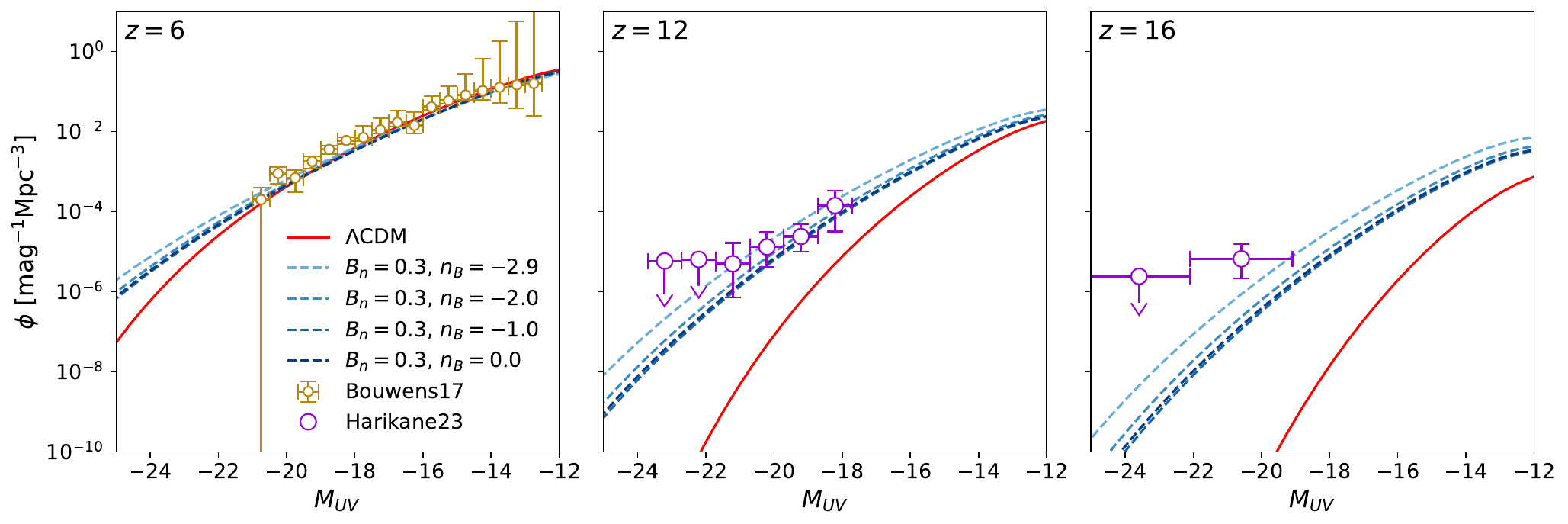}
\caption{\label{fig:LF_B03} Same as Fig. \ref{fig:LF_nm29} but for PMF Model 5--8; darker shades correspond to steeper spectra of PMF.}
\end{figure*}

The excess of bright high-$z$ galaxies is thus an imprint of the non-Gaussianity of the PMF-induced density fluctuations, and not of the additional small-scale power that PMFs generate. For the fiducial field strength the latter is invisible in the UVLF at any magnitude; it leaves a trace only for $B_n \gtrsim 1$~nG, for which the enhancement of $\sigma(M,z)$ reaches $M\sim 10^{10}\,h^{-1}\,M_\odot$, which is larger enough than $M_\mathrm{turn}$. Figures~\ref{fig:LF_nm29}--\ref{fig:LF_B03} therefore show only $M_\mathrm{UV}\le -12$, where the predictions can be confronted with observations, and sufficiently larger than turnover halo mass.

Having established that the non-Gaussianity drives the enhancement, we next examine how the UVLF depends on the PMF parameters $B_n$ and $n_B$. Figure \ref{fig:LF_nm29} shows the results of galaxy UVLFs at $z=6, 12,$ and $16$. We compare the $\Lambda$CDM model with PMF-induced models, incorporating both the enhancement of matter power spectrum and the non-Gaussianity of density fluctuations. Here, the spectral index of PMFs is fixed at $n_B=-2.9$, and we plot the cases with different field strengths, $B_n=2.0$ -- $0.1$.
The astrophysical parameters ($f_{*,10}$, $\alpha_*$, $M_\mathrm{turn}$, $t_*$) are held fixed at the fiducial values quoted above; only the HMF differs between the $\Lambda$CDM and PMF cases.
The response of the UVLF to variations of these astrophysical parameters is discussed in Appendix~\ref{app:astro-sensitivity}.
We also show the recent observational data for comparison \cite{2017ApJ...843..129B, 2023ApJS..265....5H}.

At higher redshifts ($z=12$ and $16$), the PMFs significantly increase the number density of massive halos and bright galaxies ($M_\mathrm{UV}<-18$) due to the non-Gaussian tail of one-point PDF.
James Webb Space Telescope (JWST) reported that it had observed higher number density of high-$z$ galaxies than theoretically expected~\cite{2023ApJS..265....5H}.
We find that the PMF model with the fiducial astrophysical parameters reproduces the observed bright-end abundance to within the reported uncertainties at $z=12$ and $16$ for $M_\mathrm{UV}\lesssim -18$.

This is the main qualitative point of the present work: within the limiting case of Sec.~\ref{sec:results-HMF}, a nearly scale-invariant PMF can plausibly account for the excess of bright galaxies at $z\gtrsim 12$ reported by JWST through the non-Gaussian tail of $P_\mathrm{1pt}$ alone, as shown in Fig.~\ref{fig:LF_NGcom}, without invoking additional astrophysical boosts of the star-formation efficiency.
On the other hand, at lower redshift ($z=6$), the contribution of PMFs is relatively small and all the models in this figure are consistent with observational data \cite{2017ApJ...843..129B}.

When increasing the magnetic field strength $B_n$, the faint-end galaxy abundance increases while the bright-end UVLF remains nearly constant. Our results show that PMF non-Gaussianity enhances the abundance of high-$z$ massive galaxies, and that a stronger magnetic field increases the number density of small galaxies.
This insensitivity of the bright end to $B_n$ reflects our assumption that the total density field inherits the PDF shape of the PMF-induced component; a more realistic treatment, in which the Gaussian adiabatic and the non-Gaussian PMF-induced fluctuations are combined with their respective amplitudes, would introduce an explicit $B_n$ dependence, and we leave this to future work.
As we have seen so far, the enhancement of the matter power spectrum from PMFs affects the $\sigma$--$M$ relation only at $M \lesssim 10^{10}\,h^{-1}\,M_\odot$, and the exponential tail of the one-point PDF can increase the number density of rare objects: high-redshift and massive halos and galaxies.

We also show the galaxy UVLF while fixing $B_\mathrm{n}= 0.3$ nG and varying the PMF spectral index $n_B=-2.9$ -- $0$ in Fig. \ref{fig:LF_B03}. While the UVLF with $n_B=-2.9$ is consistent with all the observation data \cite{2017ApJ...843..129B, 2023ApJS..265....5H}, the steeper case such as $n_B=0$ shows the slightly smaller than the observed data with $1 \sigma$ error at $z=16$.

\section{Conclusion}
\label{sec:conclusion}

In this work, we quantify the non-Gaussianity of matter density fluctuations sourced by PMFs and propagate it into the HMF and the galaxy UVLF within the Press--Schechter framework.
By numerically evaluating the spatial distribution of the Lorentz source term $\nabla \cdot [(\nabla \times \bm{B}) \times \bm{B}]$ for an ensemble of Gaussian realizations of the magnetic fields, we show that the one-point PDF of $\delta_M$ deviates from the Gaussian case by less than $\sim 20\%$ in the central region $|\delta_M|\lesssim 2\sigma$, but is enhanced by about three orders of magnitude in the tail at $|\delta_M|\sim 6\sigma$ (Fig.~\ref{fig:PDF_delta_PMFs}). This non-Gaussian tail is well approximated by an exponential function with transition point $\nu_\mathrm{tr}$ and the central Gaussian width $\sigma_\nu$, which are summarized in Table~\ref{tab:fitting-parameters}.
Propagating this PDF through the Press--Schechter framework as a limiting case for the total density field, we find that the abundance of massive halos and bright galaxies at high redshifts is enhanced relative to $\Lambda$CDM, providing a possible explanation for the bright-end excess of the JWST UVLF at $z\gtrsim 12$. 
Comparing the UVLF computed with the PMF-induced power spectrum alone against the one that also carries the non-Gaussian PDF (Fig.~\ref{fig:LF_NGcom}), we further find that the enhancement on the bright end comes entirely from the non-Gaussian tail. The additional small-scale power with the PMF parameter $B_\mathrm{n}= 0.3$ and $n_B=-2.9$ is confined to galaxies more than an order of magnitude lighter than ones hosted by halo mass $M_\mathrm{turn}$, which the duty cycle removes from the UVLF altogether.

Several recent studies have investigated how PMFs affect small-scale structure formation using cosmological simulations with PMF-aware initial conditions; a representative example is Ref.~\cite{2024JCAP...07..027R}. There, hydrodynamical simulations are performed in which standard $\Lambda$CDM displacements are combined with additional particle displacements sourced by the PMF Lorentz force, and the HMF and baryon fractions of high-$z$ haloes are analyzed.
In the present work, we take a complementary route.
In the linear approximation, we assemble PMF-sourced one-point statistics from 50 independent PMF realisations per parameter set (Sec.~\ref{sec:NumericalSetup}) and fit the resulting distribution.
We then propagate it through Press--Schechter to estimate how the far tail of the PDF affects the abundance of rare massive haloes and bright galaxies.
Comparing with the recent hydrodynamical simulation studies, we sacrifice detailed fidelity of small-scale nonlinear time evolution. However, we obtain a focused estimate of how PMFs affect high-$z$ massive haloes with a large statistical significance.
Large-scale and high-resolution cosmological MHD simulations with consistent initial conditions and sufficient statistical sampling could, in principle, treat both small and massive halos equally. In practice, however, such simulations face a major barrier in the form of computational cost.

More caveats of our calculation should be emphasised. First, we have assumed the critical density contrast for spherical collapse to be a constant, $\delta_c = 1.69$. Ref.~\cite{2014JCAP...08..017S} investigated the spherical collapse model in the presence of PMFs and found that the effective $\delta_c$ can be significantly larger than the $\Lambda$CDM value, particularly at high redshifts, which would reduce the high-$z$ halo abundance relative to our fixed-$\delta_c$ estimate. Our predictions in Figs.~\ref{fig:HMF}--\ref{fig:LF_B03} should therefore be regarded as an upper bound on the UVLF enhancement attributable to the non-Gaussian PDF alone.
Second, the pressure-free approximation underlying Eq.~\eqref{eq:delta_m} is valid above the magnetic Jeans length. Since the main scope of this paper is non-Gaussianity produced by PMFs and its impact on the abundance of massive halos and galaxies, this assumption does not harm our results.
Third, we have neglected magnetic dissipation heating on the IGM. This effect significantly affects the baryon physics, rather than the gravitational astrophysics treated in this paper. If we include the effects of PMFs on the baryon physics (such as star formation, reionization, AGN activities, and so on), we should perform large-scale cosmological MHD simulations.
We leave a detailed investigation of these points for future work.

\begin{acknowledgments}
This work was supported by The National Natural Science Foundation of China (NSFC) (Grant No. W2433009) and the National SKA Program of China (No. 2020SKA0110401). HS was also supported by NSFC (Grant No. 12103044), Yunnan Provincial Key Laboratory of Survey Science with project (No. 202449CE340002) and Yunnan Revitalization Talents Support Plan. SS was supported by the Japan Society for the Promotion of Science KAKENHI Grant Numbers JP24K17043 and 26H02044~(SS), and also acknowledges support from DAIKO FOUNDATION and Hirose Foundation.
We thank Yu'ichi Harikane for discussion about the JWST UVLF.
We have developed the calculation code to generate magnetic fields with Kiyotomo Ichiki.
The authors acknowledge the Tsinghua Astrophysics High-Performance Computing platform at Tsinghua University for providing computational and data storage resources that have contributed to the research results reported within this paper.

\end{acknowledgments}

\appendix

\section{Effects of astrophysical parameters on the UVLF}
\label{app:astro-sensitivity}

In the main text, the four astrophysical parameters entering Eqs.~\eqref{eq:UVLF}--\eqref{eq:UVmag}, namely $f_{*,10}$, $\alpha_*$, $M_\mathrm{turn}$, and $t_*$, are held fixed at the fiducial values of Ref.~\cite{2019MNRAS.484..933P}.
Here we briefly examine how the $\Lambda$CDM UVLF responds when each parameter is varied around that fiducial choice, with the remaining three astrophysical parameters held fixed and with no PMF contribution.

\begin{figure*}
\centering
% Renamed to match the figure number in print (was Fig06_LF_astro_sensitivity.pdf).
\includegraphics[width=1.0\linewidth]{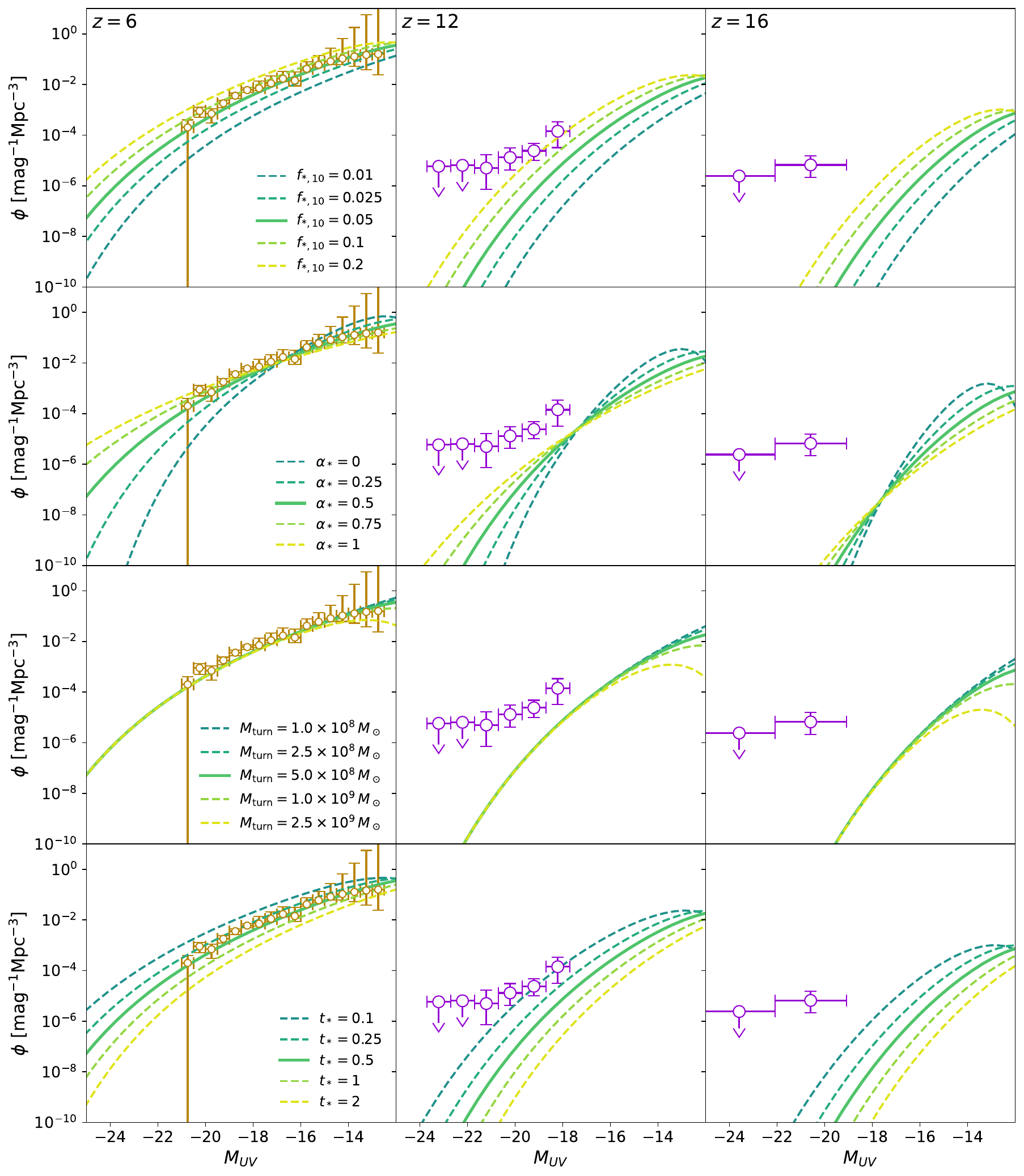}
\caption{\label{fig:LF_astro} Sensitivity of the $\Lambda$CDM UVLF to the astrophysical parameters.
From top to bottom, each row varies $f_{*,10}$, $\alpha_*$, $M_\mathrm{turn}$, or $t_*$ around the fiducial values, with the other three parameters held fixed; the thick solid line in every panel is the fiducial choice.
From left to right, the columns correspond to $z=6$, $12$, and $16$.
Observational data points are the same as in Fig.~\ref{fig:LF_NGcom}.
}
\end{figure*}

Figure~\ref{fig:LF_astro} shows the resulting UVLFs at $z=6$, $12$, and $16$.
Varying $f_{*,10}$ or $t_*$ mainly shifts the bright-end abundance, whereas $M_\mathrm{turn}$ affects the faint end alone, leaving the bright end untouched.
The slope $\alpha_*$ acts differently from either: it tilts the UVLF about the pivot mass $M=10^{10}\,M_\odot$ of the stellar-to-halo mass ratio, so that the bright and faint ends respond in opposite senses.
Within this astrophysical parameter space, however, no single choice simultaneously reproduces the observed UVLFs at all three redshifts: parameters that raise the bright-end abundance at $z=12$ and $16$ tend to overproduce galaxies at $z=6$, and vice versa.
By contrast, as shown in Sec.~\ref{sec:HMF-UVLF}, PMF-induced non-Gaussian tail of $P_\mathrm{1pt}$ can increase the high-$z$ bright-end abundance while remaining consistent with the $z=6$ data for the same fixed astrophysical parameters.

\sloppy
\bibliography{apssamp}% Produces the bibliography via BibTeX.

\clearpage

\end{document}